\documentclass[letterpaper, 10 pt, conference]{ieeeconf}

\IEEEoverridecommandlockouts                            

\usepackage[utf8]{inputenc} 
\usepackage[T1]{fontenc}    
\usepackage{hyperref}       
\usepackage{url}            
\usepackage{booktabs}       
\usepackage{amsfonts}       
\usepackage{nicefrac}       
\usepackage{microtype}      
\usepackage{lipsum}
\usepackage[T1]{fontenc}

\usepackage{cite}

\usepackage{amsmath,amssymb}
\usepackage{algorithmic}
\usepackage{graphicx}
\usepackage{subfigure}
\usepackage{textcomp}
\usepackage{gensymb}
\usepackage{epsfig} 
\usepackage{xcolor}

\usepackage{booktabs}
\usepackage{array}
\usepackage{multirow}

\title{\LARGE \bf Vibration Improves Performance in Granular Jamming Grippers}

\author{Raghav Mishra$^{1,2}$, Tyson Philips$^{2}$, Gary W. Delaney$^{1}$, and David Howard$^{1}$

\thanks{$^{1}$ Data61, CSIRO, Australia; contact {david.howard@csiro.au}}
\thanks{$^{2}$ School of Mechanical and Mining Engineering, University of Queensland, Australia; contact {r.mishra@uqconnect.edu.au}}

}

\begin{document}
\maketitle
\thispagestyle{empty}
\pagestyle{empty}

\begin{abstract}
Granular jamming is a popular soft robotics technology that has seen recent widespread applications including industrial gripping, surgical robotics and haptics. However, to date the field has not fully exploited the fundamental science of the jamming phase transition, which has been rigorously studied in the field of statistical and condensed matter physics. This work introduces vibration as a means to improve the properties of granular jamming grippers through vibratory fluidisation and the exploitation of resonant modes within the granular material. We show that vibration in soft jamming grippers can improve holding strength, reduce the downwards force needed for the gripping action, and lead to a simplified setup where the second air pump, generally used for unjamming, could be removed.  In a series of studies, we show that frequency and amplitude of the waveforms are key determinants to performance, and that jamming performance is also dependent on temporal properties of the induced waveform.  We hope to encourage further study in transitioning fundamental jamming mechanisms into a soft robotics context to improve performance and increase diversity of applications for granular jamming grippers.

\end{abstract}


\section{Introduction}

In the last decade, robotics has evolved beyond its traditionally rigid roots through the paradigm of soft robotics. Soft robots take advantage of soft and compliant structures, exploiting the mechanical properties and non-linear behaviours of their constituent materials to provide rich, adaptive behaviours.  Nowhere is this more prevalent than in the tasks of gripping and manipulation \cite{shintake2018soft}, where compliance and variable stiffness are harnessed to conform to irregular object shapes, maintaining larger contact surfaces and exerting fewer extraneous contact forces. 

Many soft robot structures are pneumatically powered due to the simplicity and low cost of the actuation. However, they often have issues with insufficiently large force exertion capability and low resistance to deformation, which limits potential applications. The most common response to these problems are variable-stiffness soft structures which allow the robot's compliance to adapt as desired to the situation. A popular implementation of a variable stiffness soft structure exploits the jamming phase transition, through e.g., laminar, fibre, or granular jamming \cite{fitzgerald_review_2020}. Granular jamming is the most common of the three, whereby granular materials are driven from fluid-like to solid-like states through the variation of (mainly frictional) inter-particle forces under the presence of internal stress. Granular jamming provides a simple soft structure that exhibits a wide range in possible stiffness values, rapid stiffness variation, and a wide array of possible shapes and sizes.

\begin{figure}[t!]
\centering
\includegraphics[width=0.95\columnwidth]{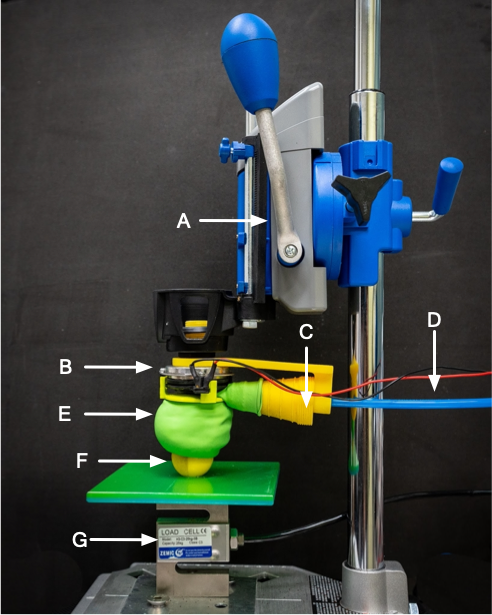}
\caption{Vibration-based fluidizing jamming gripper setup. (A) linear drill press stand (B) vibration speaker (C) 3D printed unit/membrane mount (D) silicone tube (to vacuum pump) and electrical wiring (to vibration unit) (E) latex membrane and coffee ground gripper (F) 3D printed target object (G) Load cell.}
\label{fig:fig1}
\end{figure} 

The archetypal example of granular jamming for robotics is the granular jamming-based universal gripper \cite{brown_universal_2010}. A  granular material (e.g., ground coffee) is surrounded by a soft membrane and jammed by application of negative-pressure from a vacuum pump after being pushed down and moulded around a target object that is to be picked up. The resultant stress from the atmospheric pressure causes a jamming phase transition to occur; the gripper hardens and is able to maintain a grip via the interaction between the target object and the granular material contained within the soft membrane. This design allows for robust 'universal' gripping that is suitable for a variety of objects and does not require software to solve complex manipulation planning problems, simultaneously offering a simple, generally-applicable gripping solution. 

Although granular jamming has been explored in a variety of configurations and applications in robotics, research in to the effects of the underlying material properties and related granular phenomena is limited. In this paper, we explore the possibility of enhancing jamming grippers through induced vibration, primarily seeking to cause granular fluidisation (e.g., \cite{kapadia_design_2012}).  We develop a novel setup where a small, versatile vibration element is mounted on a 3D printed bracket and placed in direct contact with the latex membrane surrounding the granular material.  We show that vibration has the ability to improve holding force, reduce the pre-loading forces exerted on the target object, and reduce the mechanical complexity of grippers. As our vibration unit can be simply programmed, we are additionally able to show that different input vibration waveforms can result in different holding forces, which provides further opportunity in optimising the gripper for particular objects and applications.

\section{Theory of operation}
In physics, there are various perspectives from which one can consider granular materials \cite{baule2018a}. The jamming behaviour of granular materials is often explained through a thermodynamic analogy, wherein the external forces acting on the jammed material are similar to atmospheric pressure exerted on normal materials, which cause the majority of materials to condense and solidify. Temperature also plays a critical role in determining the phases of matter. Granular materials behave analogously to this with the kinetic energy in the system (such as that imparted by vibration) representing the `temperature'. Therefore, one can in certain circumstances counteract the jamming effects that may occur in a granular material by producing vibration within the system, thus increasing the `temperature'. \cite{hao_defining_2018} proposed an elegant and simple definition of `temperature', $T$, for granular materials as 
\begin{gather}
\frac{3}{2}k_B T = \frac12 m\left\langle v\right\rangle^2
\end{gather}

In this case, $k_B$ is an equivalent to the usual Boltzmann constant from statistical mechanics, and $\left\langle v\right\rangle$ is the root-mean-square velocity of particles. However, there are multiple other temperature-like parameters that have been proposed for a vibrational input with amplitude $A$, angular frequency $\omega$, particle diameter $d$ and gravity $g$, such as the dimensionless peak acceleration, $\Gamma = \frac{\omega^2 A}{g}$, dimensionless vibration velocity $v = \omega A$ and the shaking strength parameter $S = \frac{\omega^2 A^2}{gd}$. Regardless of the exact parameter being introduced, many papers have been successful at expressing properties of jamming in terms of these simplified temperature-like parameters, although with limitations \cite{windows-yule_resonance_2015}.  

Vibration of granular materials has been extensively studied and applied in industrial contexts for unjamming, in particular for industries that require flow of granular materials such as in food processing, pharmaceuticals, and agriculture \cite{delaney2012}. However, there is very little work that explores vibration for jamming grippers, where the tendency of a jammed granular material to fluidise in the presence of vibration can be used to counteract undesired jamming. A key drawback in traditional granular jamming grippers is that partial jamming occurs during the object moulding processes.  As the gripper pushes down on the object and begins to conform to the shape,  the grains transmit forces from the gripper to the object. This transmission causes stress to build up within the gripper, causing jamming which reduces the gripper's compliance, thus producing a detrimental effect on performance.  A degree of internal pressure within the gripper is required for optimal moulding around the target object, which we propose to cause via a vibration element attached to the gripper mount.  Vibration increases both compliance and the contact surface area during the moulding process, leading to greater holding forces. Additionally, this produces reduced forces on the test object as it allows the gripper to relax the stress built up in the grains during the moulding process. This is beneficial for softer and more fragile objects where a more delicate touch is required to prevent deformation, damage or movement to an object.

\section{Background}
\subsection{Soft Gripping}

The field of gripping and manipulation has a rich research literature exploring soft structures and spans a variety of distinct design approaches including multi-fingered soft grippers \cite{yang_hybrid_2020, li_passive_2017}, 
programmed deformation in bag-style membranes \cite{howard_one-shot_2021}, origami-style grippers \cite{yang_grasping_2021}, and
 bio-inspired grippers taking inspiration from diverse sources including tentacles  \cite{martinez_robotic_2013} and chameleon tongues \cite{noauthor_adaptive_nodate}.

\subsection{Granular Jamming in Soft Robots}
Since the introduction of the original jamming gripper, vacuum-actuated granular jamming has been used in a variety of applications.  Furthermore, they have been used in numerous configurations, including more complex multi-finger designs and in combination with traditional pneumatic structures.

The gripper displayed in this paper is a modification of the original universal jamming gripper \cite{brown_universal_2010}. Previous work has been performed on modifying the original design to push the grippers capabilities, such as adding positive pressure to counter the inability to grip traditionally difficult objects \cite{amend_positive_2012} and exploring force-feedback for a closed loop approach \cite{nishida_development_2014}. The exploration of these grippers capabilities have also been applied to different environments including underwater usage \cite{licht_universal_2016}, and a scaled-up variant used on a crane \cite{miettinen_granular_2019}.

Materials have been explored as a means to tune jamming performance, with both shape and size of grains playing a major role \cite{howard2021shape}, as well as multi-material membranes to induce programmed deformations \cite{howard_one-shot_2021}.  Evolutionary algorithms \cite{fitzgerald_evolving_2021} were used to generate bespoke grains for soft jamming grippers, as well as to explore the possible configuration space of granular morphologies in order to optimise properties for bespoke granular design \cite{delaney_multi-objective_2020, delaney_utilising_2019}.  The exploitation of granular jamming's variable stiffness properties have been explored in a plethora of morphologies beyond 'bag-style' universal gripper designs, including multi-finger grippers, paws for legged robots \cite{chopra_granular_2020,howard_one-shot_2021}, prostheses \cite{chen2019learning} and haptic feedback devices \cite{zubrycki_novel_2017}. Aside from stiffness variability, there has been little work in exploring the utilisation of other granular physics effects in jamming-based robots.

Fluidisation has been covered in a few studies, e.g., through periodically \cite{amend_positive_2012} or continually modulating the air pressure \cite{kapadia_design_2012} inside the membrane.  These setups require two vacuum pumps or elaborate pneumatic circuits to function, and do not exploit vibration.  

Vibration has been briefly mentioned \cite{amend_soft_2016}, where a vibration motor is used in place of a positive pressure pump to unjam granular material after a grip is executed, rather than as an active element during a grip as in our case.  It was found that smaller vibration motors which could be embedded in the granular material did not have enough power output to unjam the gripper, while larger motors which could sustain the power could not be embedded within the gripper without changing bulk properties.   We use a compact, powerful audio exciter as a vibration element which is mounted outside the gripper membrane. This choice also allows more flexibility in vibrational waveform inputs due the actuator's large bandwidth, unlike vibration motors which need to be tuned to desired frequencies. 

Compared to the above work, our method of fluidisation by vibration is superior in many regards.  Mainly, our fluidisation is decoupled from application of vacuum, and is programmable with a wide range of input waveforms.  Importantly, ours is the first work to exploit vibration during a grip.  Our method is mechanically simpler as it does not rapidly alternate pump flow through a valve, which requires 2-way valves and a controller or two pumps, and causes extra fatigue cycles in the gripper and supporting equipment. Vibration is also amenable to analysis of its physics and has a richer literature of work for studying its theoretical behaviour, as discussed in Section \ref{vibra}.  

\subsection{Vibration in Jamming Systems}
\label{vibra}
Vibratory Fluidised Beds are commonly used for unjamming in industries dealing with food, powder, pharmaceutical and agricultural processing, e.g., to enable smooth flow down a hopper \cite{janda_unjamming_2009}. 

Much work has been done trying to understand the origins of the properties of granular materials on the mesoscopic scale, using  created 2D lattice gas automata models for studying vibration in granular materials \cite{nicodemi_compaction_1997, nicodemi_frustration_1997}, showing properties such as glass transition and density relaxation.   Granular materials vibrated using vibrational beds show different internal energies depending on the input frequencies, even if the energy corresponding to the vibrational system is kept constant, due to the resonant modes of the bulk material\cite{windows-yule_resonance_2015}.  Discrete Element Method (DEM) simulations of granular matter under vibration showed the presence of linear vibrational modes at low amplitudes with modulus softening behaviour, which was attributed to the remaking of internal force chain contacts during vibration \cite{lemrich_dynamic_2017}.

\subsection{Motivation}
Overall, placing our work in the context of the wider literature allows us to delineate several novel contributions.  We present the first jamming gripper that leverages vibration as a means of active fluidisation, e.g., during a grip operation. with programmable output.  Vibration is shown to increase grip strength, reduce stress during object moulding, and increase surface contacts between gripper and object.  We use our flexible audio exciter-based setup to run experiments with a variety of input waveforms. We also delve further into exploiting granular matter physics theory to motivate our experimentation.


\section{Materials and Design}
The proposed vibration-enhanced gripper is similar to a traditional granular jamming gripper \cite{brown_universal_2010}, with the addition of a vibration element attached  from above to the soft membrane. The vibration element vibrates to fluidise the granular material in the gripper. This fluidised gripper is pushed onto the target object to mould around its shape before being jammed using the connected vacuum pump.  

A Qualatex latex balloon was used for the soft membrane and filled with ground coffee grains. A frame for the gripper to hold the vibration exciter was 3D printed using a Stratasys Object500 Connex3 printer.  The balloon membrane neck and lip can not point upwards as they would in a traditional jamming gripper due to the position of the vibration element. As such, the neck is reoriented to feed out from the side from which it can access the vacuum pump tube. It was observed that not all parts of the main body of a balloon are consistent in their ability to act as a gripping surface. The tip or `drip point' of the balloon acts as the most consistent gripping surface. While the sides of the body create a depression when pushed against the object, the balloon is not able to well envelop the object. Therefore the top of the balloon is angled to point down while the neck and lip escape from the side of the vibration element into the frame.

For our application, we must vibrate the gripper with enough power to cause fluidisation of the granular material, without hampering either its bulk properties or its mechanical compliance. Additionally, it is desirable if the vibration element allows a large range of possible input waveforms with additional control over frequency of oscillation. This is not trivial as many types of vibrating mechanical systems, such as those made of asymmetrical masses attached to motors, are often heavily reliant on their natural frequencies of vibration and have little control over frequency. These sets of requirements made audio speaker exciters an ideal choice, having a small footprint and large bandwidth to accommodate much of the human hearing range from 20Hz to 20kHz within its pass-band. We chose to use the Tectonic 30W TEAX32C30 Speaker-based exciter drivers, which were found to output enough power to cause fluidisation, while having a suitable bandwidth.

\section{Experimental setup}

\begin{figure}[h!]
\centering
\includegraphics[width=0.95\columnwidth]{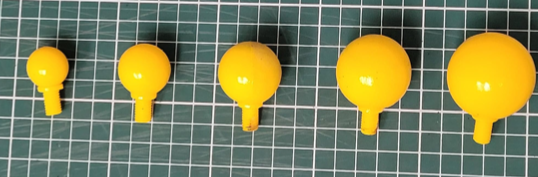}
\caption{The range of 3D printed sphere objects used for testing. Sizes range from 20mm to 40mm diameter, in range increments of 5mm.}
\label{fig:fig3}
\end{figure}

Spherical target objects of sizes from 20mm to 40mm in diameter were 3D printed to test the holding strength of our gripper (Fig.\ref{fig:fig3}). These sizes were chosen as the gripper diameter was approximately 45mm and traditional jamming grippers are known to display degraded performance for objects larger than about half the size of the gripper \cite{brown_universal_2010}. This provided a large range of testing possibilities that are traditionally considered on the edge of or beyond the capability of our universal gripper.

\begin{figure}[h!]
\centering
\includegraphics[width=0.95\columnwidth]{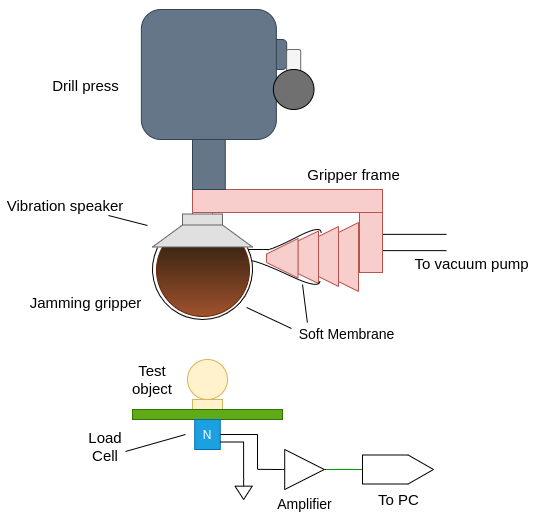}
\caption{Simplified diagram of the test setup used for the experiments in the paper, showing the gripper connected to the drill press and the test object for measuring holding forces}
\label{fig:fig2a}
\end{figure}

The object to be tested was attached to a Zemic H3-C3-25kg-3B load cell, which was connected to a Raspberry Pi through an amplifier.  The jamming gripper was attached to the drill press via a printed screw thread. During a test, data was sequentially recorded through the Raspberry Pi. The gripper was lowered to a set height and pressed against the target object, then jammed using the vacuum pump.  The drill press was slowly returned to its starting position to measure the holding force (Fig.\ref{fig:fig2a}). The point at which the upwards force from the gripper dropped from its peak to zero due to slipping was the holding force.

Two experiments were carried out using the gripper. The first experiment aimed to understand the effects of vibration on jamming gripper performance. This included the improvement in holding force, the decrease in the downwards pre-loading force needed to achieve a grip, and the effect of the vibration frequency on the holding force for a constant frequency. The second experiment used a variety of temporal waveforms the gripper as a way to explore effects of time-varying inputs.

\section{Results}

\subsection{Experiment 1: Effect of constant vibration}
\begin{figure}[h!]
\centering
\includegraphics[width=0.95\columnwidth]{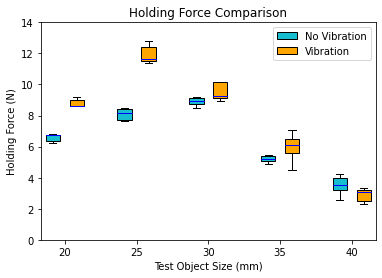}
\caption{Box and whisker plot displaying holding force variation with test object size for a constant vibration at 250Hz, showing a significant increase in grip strength for all but the largest sized object.}
\label{fig:fig4}
\end{figure} 

In the first experiment, the gripper height was adjusted to be 1 cm above the top of the object, and pushed down all the way through the drill press, or until prevented from further movement by resistance of the object. The vibration element was turned on with constant 250Hz sinusoidal vibration for 5 seconds as downwards pressure was maintained on the object.  Then when the vibration was turned off, the vacuum pump attached to the gripper was turned on and the gripper was pulled back up until the gripped object slipped out. The force on the gripper was continuously measured through the load cell. Five repeats of this were run for each test object and the process was repeated without vibration to isolate the effects of vibration on comparative holding strength.

 Results show a distinct improvement in average holding forces for all object sizes up except 40mm. The amount of improvement varies between sizes with a dramatic 50\% increase in holding force with the 25mm object. Peak force similarly improves with vibration in all cases except 40mm.  The most marked differences occur with 20mm (6.6N vs. 8.7N average holding force) and 25mm (8.2N vs 11.9N) objects, in the latter case representing a 145$\%$ improvement with vibration.

\begin{figure}[h!]
\centering
\includegraphics[width=0.95\columnwidth]{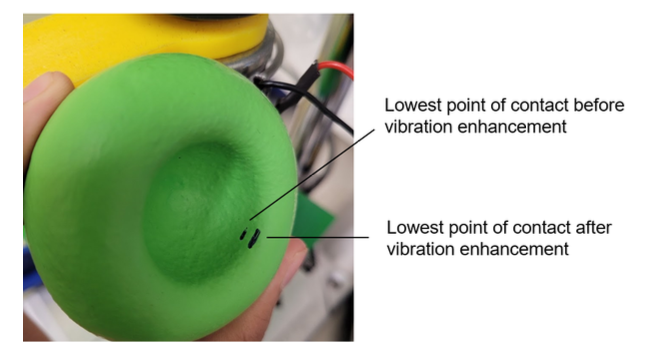}
\caption{Vibration applied to a jamming gripper on 25mm sphere object achieves a lower point of contact when gripped, leading to a larger contact area for gripping}
\label{fig:fig5}
\end{figure}

We can explain these results in terms of improved gripper-object contact.  We repeat the gripping test for the 30mm object (the centre of the object size range). We mark the edge of the object impression with a marker after the gripper comes to a stop before vibration and after the gripper has experienced vibration. As shown in Figure \ref{fig:fig5}, the marks are $\approx$2.5mm apart, indicating an improved contact angle and increased enveloping of the object. This improved contact angle results in an approximately O-ring strip of additional contact area, providing a larger area for increased grain-object contacts and augmented static friction effects. This explanation also accounts for relatively smaller improvement in holding forces for larger test objects. A larger gripper would have a smaller contact angle, $\theta$, as per \cite{brown_universal_2010}, and the vertical component of the force provided by the additional O-ring shaped area is less for a smaller contact angle.  

\begin{figure}[h!]
\centering
\includegraphics[width=0.95\columnwidth]{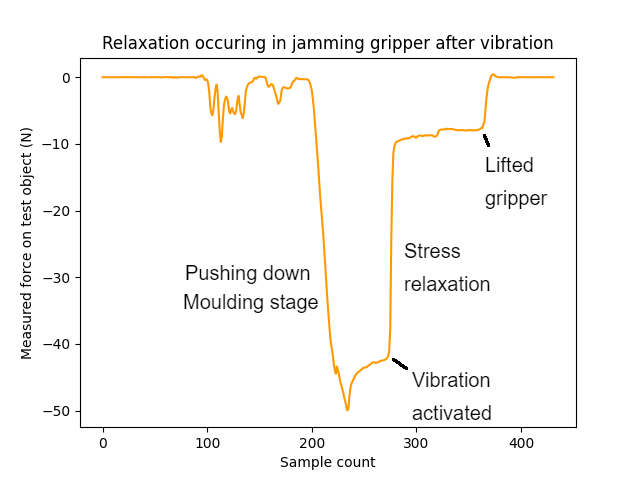}
\caption{Stress relaxation within the jamming gripper occurs when vibration is turned on, leading to a decrease in the downwards force applied to the object.}
\label{fig:fig6}
\end{figure} 

We assess the impact of vibrational fluidisation on the downwards force required to mould around the object. We run a test where the gripper is pushed down to mould around the 30mm sphere object. The vibration element is turned on while the position of the gripper is kept fixed. These vibrations result in rapid stress relaxation within the gripper as the particles rearrange to pack better around the target object (Fig.\ref{fig:fig6}), leading to a rearrangement of the force network within the granular material and an overall reduction in downwards force on the target object of $\approx$80\%.  

It is generally useful to minimise the force required to grip an object, but is especially desirable when the object is soft or fragile as it would cause less deformation or fewer forces to contribute to fracture respectively. Note that for Fig.ref{fig:fig6}, the vacuum pump was not turned on for the gripping action as it is not necessary to show the stress relaxation.

Finally, we quantify the effect of vibration frequency on gripping behaviour.  We conduct an experiment comparing the result of the holding strength test used before across different constant input vibration frequencies. Object size is kept constant at 30mm. The frequency is varied first from 100Hz to 700Hz in intervals of 100Hz. As the change plateaus, we subsequently test large increases in frequency between 3kHz and 5kHz.

\begin{figure}[h!]
\centering
\includegraphics[width=0.95\columnwidth]{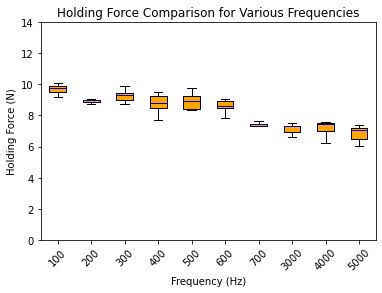}
\caption{The holding force from gripper decreases as the vibrational frequency increases}
\label{fig:fig7}
\end{figure} 

A clear trend in holding force is observed, starting from around 10N  at 100Hz and decreasing to around 7N from around 700Hz, around which the holding force plateaus for higher frequencies. We propose that this is due to the jamming gripper having low frequency fundamental modes which are not as activated by higher frequencies. Additionally, it is well known that lower frequency modes within granular materials have higher participation ratios i.e. the modes involve more global involvement of the grains in vibration, while higher frequency modes dissipate energy mostly into the part of the material that is close to the vibration element \cite{chen_low-frequency_2010}.  Results indicate that a lower frequency is beneficial, as the higher participation would imply that force chains in the lower half of the gripper are broken away from the vibration element more easily.  This also opens up the possibility to target higher frequency localised vibrations into specific regions of a jamming structure, however such experimentation is out of scope for our current study.

\subsection{Experiment 2: Effect of temporal vibration patterns}

Our second experiment aimed to observe if there are any effects on the holding forces as a result of time-dependent frequency and amplitude variations in the input waveform. These time dependent effects would affect the behaviour as high amounts of vibration allow more random movement and increased rearrangements of force chains.  Conversely low vibrations allow grains more time to relax by providing them sufficient energy and time to explore lower energy configurations. Finally, a lack of vibrations leads to freezing of the current state of the material due to the inherently high damping ratios in granular materials.  We can analyse the vibrational response of our granular material in terms of the modes that are activated. Like all structures, granular structures have natural vibrational modes. Vibratory inputs of different frequencies will activate different modes which can vary in their properties and cause local energetic interaction variations over the volume of the gripper such as the previously-described localisation of high frequency modes.

\begin{figure}[h!]
\centering
\includegraphics[width=0.95\columnwidth]{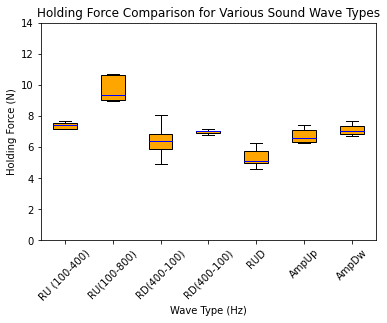}
\caption{Change in holding force as the input waveform changes. RU and RD are ramps up and down in frequency. RUD is a ramp up followed by a ramp down in frequency. AmpUp and AmpDw are ramps up and down in amplitude.}
\label{fig:fig8}
\end{figure}

A myriad of experiments were conducted with varying vibration input waveforms in the form of ramps up and down in frequency (chirps) and amplitude. Specifically, the input frequency was ramped up, ramped down, and ramped up then down (RU, RD, RUD respectively, see Fig.\ref{fig:fig8}) from 100 to 400Hz. The first two of those were repeated for a range of 100Hz to 800Hz. Lastly, the frequency was held constant at 250Hz while the amplitude was either ramped up from 0 to max amplitude (full volume), or ramped down from max amplitude to 0. For consistency, ramps in frequency were all carried out at 30Hz/s.

Results demonstrate a clear dependency between temporal properties of the input signal and gripper behaviour.  Frequency ramps are seen to have more of an effect than amplitude ramps, and specifically RU(100-800) displays higher holding force than a constant 250Hz vibration. In some cases the result is easily explained, as slowly ramping down from high vibration to low vibration allows the granular material to settle into a more compact and stable configuration in a process similar to annealing. It is interesting to note that the results show not only effects on performance but also the variance of the achieved holding forces. 

However, (i) the performance of constant 250Hz vibration sits in between RU(100-800) than RU(100,400), and (ii) up-down ramps show different performance compared to pure ramp downs, indicating a more complex, unintuitive underlying mechanism relating waveform to grip strength, and suggesting further experimentation is necessary to explore the optimal waveform for a particular application.

\section{Conclusion}

Despite the extensive exploration of granular jamming-based universal grippers in a variety of applications, robotics literature has largely failed to exploit many of the results and theory that exist in the fields of condensed matter physics that underlie granular materials. In this work, we proposed a novel jamming gripper design informed by this theory that uses compact, flexible vibration elements to fluidise the jamming gripper during a grip. An audio speaker exciter was attached to the membrane of the jamming gripper from above and transmits vibrations through the membrane to the granular material. The resulting gripper maintained higher holding forces on spherical test objects, presented larger contact surface area between object and gripper, and produces significantly less downwards force as it moulds around the target object. We tested the gripper with a variety of input frequencies, showing the effect of frequency on grip strength that favours lower frequencies. More complex chirp and amplitude ramp waveforms strongly influence gripper behaviour.

While the relationship between frequency and holding force was consistent and predictable, the relationships observed for the more complex input waveforms were not. Further work is needed to understand how the time-varying vibration parameters affect the granular material inside the gripper and its resultant holding force, especially for complex waveforms\footnote{We performed some initial work on more complex waveforms, showing the gripping force obtained from 5s snippets of the chorus of popular songs using the 30mm object and the test setup described in the paper. Bob Marley - 'We Jammin' = 8.91N, Beach Boys - 'Good Vibrations' = 9.03N}.  

There are significant opportunities for future research in bringing in other granular matter physics effects and theory and using them to improve jamming grippers or create other novel jamming structures. Possible avenues may include alternative ways of jamming; studying granular flow and wave propagation effects; exploring the relationship between the size of the object, frequency of oscillation and the size of the gripper; and exploiting crystallisation and compaction effect, in particular for granular materials with specified particle morphologies \cite{delaney2010g}. Some improvements are predictable using analogies to other physical materials, such as the behaviour of the amplitude ramp waveforms.  However many effects will require experimentation to determine, and have the potential to generate significant theoretical as well as practical contributions.

This is an exciting new direction in robotics research, with excellent possibilities offered from the deeper exploration of the unique emergent bulk behaviour of granular matter that is driven by complex and highly non-linear contact phenomenon on the mesoscopic scale. The field offers a rich vein of opportunities for both fundamental and applied research. We hope to encourage others to consider these possibilities and explore how they may improve or introduce new granular matter-based soft structures for robotics. 

\bibliographystyle{IEEEtran}
\bibliography{template}
\end{document}